# A Universal Multi-Hierarchy Figure-of-Merit for On-Chip Computing and Communications

Shuai Sun[1], Vikram K. Narayana[1], Armin Mehrabian[1], Tarek El-Ghazawi[1], Volker J. Sorger[1]

[1]Department of Electrical and Computer Engineering, George Washington University
800 21st Science & Engineering Hall, Washington, DC 20052, USA
Email: sorger@gwu.edu

*Abstract*—**Continuing demands for increased compute efficiency and communication bandwidth have led to the development of novel interconnect technologies with the potential to outperform conventional electrical interconnects. With a plurality of interconnect technologies to include electronics, photonics, plasmonics, and hybrids thereof, the simple approach of counting on-chip devices to capture performance is insufficient. While some efforts have been made to capture the performance evolution more accurately, they eventually deviate from the observed development pace. Thus, a holistic figure of merit (FOM) is needed to adequately compare these recent technology paradigms. Here we introduce the Capability-to-Latency-Energy-Amount-Resistance (CLEAR) FOM derived from device and link performance criteria of both active optoelectronic devices and passive components alike. As such CLEAR incorporates communication delay, energy efficiency, on-chip scaling and economic cost. We show that CLEAR accurately describes compute development including most recent machines. Since this FOM is derived bottom-up, we demonstrate remarkable adaptability to applications ranging from device-level to network and system-level. Applying CLEAR to benchmark device, link, and network performance against fundamental physical compute and communication limits shows that photonics is competitive even for fractions of the die-size, thus making a case for on-chip optical interconnects.**

## I. INTRODUCTION

The observed pace of the semiconductor industry is notably slowing down especially since the 14 nm technology node. This is driven by physical limitations relating to leakage current, thermal dissipation density, and fabrication process control (e.g. gate-oxide thickness control) becoming non-circumventable. As a compromise, 'dark silicon', which represents the part of the powered off on-chip circuit that restricted by the power and thermal budget, emerges to both emergence of extreme scaled devices and the adoption of parallelism in computing, i.e. multi-core processing [1-3]. These front-end challenges ripple through to data-communication back-end; for instance the electrical capacitance of a piece of electrical wire in 14 nm technology node is 1.65 pF/cm. Thus over 800fJ of power is dissipated to charge a 1 cm metallic wire given a VDD = 1V. With rapidly rising machine performance the communicate–to-compute overhead is increasing, making a case to use the bosonic nature of photons to enter a flatter scaling regime for communication technologies [4]. That is, Silicon photonics and possibly plasmonics may be integrated on-chip while mitigating challenges with both power density and heat dissipation problems, while extending data bandwidth. The synergies with CMOS processing, high optical index, low extinction coefficient at near infrared (NIR) frequencies, and parallelism strategies such as Wavelength Division Multiplexing (WDM) make Silicon photonics a promising option for chip and board-level communication. However, the physical size of both the optical mode and opto-electronic device footprint restricts integration density and energy-per-bit efficiency. Modal leakage, for instance, leads to crosstalk, which require waveguide-to-waveguide spacings of several micrometers [5-6]. As such the often-stated diffraction limit of light (DLL) is not so much of a limit by itself since the high refractive index of semiconductors reduces the modal cutoff of a waveguide to about 200 nm at NIR wavelengths. In fact, it is not obvious why the operating wavelength on-chip ought to be a telecommunication frequency; visible frequencies are conceivable for intra-chip applications to operate on Silicon nitride on-insulator substrates reducing the DLL to <100 nm, which is smaller than the width of a modern transistor. The actual challenge of photonics is the fundamentally weak light-matter-interaction (LMI) originating from the small dipole moment of the optical wave acting at the matter atom, which leads to 10-100's of micrometer long interaction lengths for optoelectronic devices. However, making the photon more polaritonic (matter-like), such as in plasmonics, enables strong LMIs and hence short devices which has positive effects on the device performance of the device [7-23]. Positive effects of wavelength-scale active opto-electronics are a) low electrical capacitance, b) short photon lifetimes allowing rapid re-excitation of the device (e.g. modulation, small-signal gain modulation of lasers), and high energy efficiency due to the small capacitance and voltage enabling dense and high-performing devices. Nevertheless, a polaritonic waveguide has naturally high optical losses limiting signal propagation to less than hundred micrometers. Thus, by combining the low propagation loss silicon photonic links with ultra-fast plasmonic active devices, a hybrid interconnect is able to combine high LMI active optoelectronics with low-loss passive photonic elements therefore enabling high-performance hybrid photonic plasmonic interconnects [5].

In this work, we introduce a universal FOM **C**apability-to-**L**atency-**E**nergy-**A**mount-**R**esistance (CLEAR). This FOM covers both physical and economic factors related to the evolution rate of different technology options among multiple hardware hierarchy levels; from the device building block level, over interconnect link level, to the network compute system level. By comparing the FOM value at different interconnect lengths, CLEAR is able to select the best technology option and achieve application-driven dynamic reconfigurability if the network offers the built-in overhead to do so. As such CLEAR can be regarded as a universal guideline for emerging technology options in on-chip computing and communications since it incorporates fundamental device performance and economic models. The rest of this paper is structured into three major parts as follows: 1) establish a multi-factor FOM that is able to track the compute system evolution more accurately than conventional FOMs; 2) breakdown CLEAR into the energy efficiency and the computational efficiency to show the compute system evolution in these two aspects; 3) expand CLEAR into device, link and network levels for performance comparison due to its ability to track the actual performance evolution accurately.

## II. RULES OF THE SEMICONDUCTORS

Moore's Law has been taken as the 'golden rule' of the semiconductor industry. However, with emerging technologies such as silicon photonics and plasmonics, simply counting the number of components on-chip, as a stand-alone metric does not accurately reflect the actual performance evolution (Fig. 1). Indeed the International Technology Roadmap for Semiconductors (ITRS) adjusted its predictions several times in an attempt to match development pace with the observed actual trend, which shows a decelerating rate. In fact, recent reports hypothesis that transistor scaling might stop sooner than originally anticipated [24]. Notably, Moore's Law itself was amended several times during the past decades in order to fit in the actual evolution rate [25]; originally it counted the number of components of an integrated circuit, then it shifted to a transistor size and speed scaling dominated model, and after the clock frequency saturated around 2006 driven by the current leakage and heat dissipation, parallel heterogeneous architectures emerged. As such, the original doubling rate of 12-month shifted to every 18, then 24-months. As such Dennard Scaling and the Koomey's Law introduced single factor based development models to include power density and the computation efficiency respectively to evaluate performance [26-27]. Despite accurate compute performance tracking for the 1950's and 60's, the Koomey's Law metric eventually deviates when driving factors become either complex or obsolete. Furthermore, Makimoto's Wave tracks the periodic variations between standardization and customization, and uses a four-factor FOM to quantify the evolution of the computer towards the popularization; performance (MIPS) divided by cost (power, volume and price) [28]. However, it also deviates after tracking its original growth rate for the first few decades. The deviation is attributed to the slow saturation of the clock speed, power density scaling, and the emergence of the multi-core parallel computing. Common to all four FOMs is the eventual deviation from the actual development rate, due to the limited number of factors considered during the technology evolution [29].

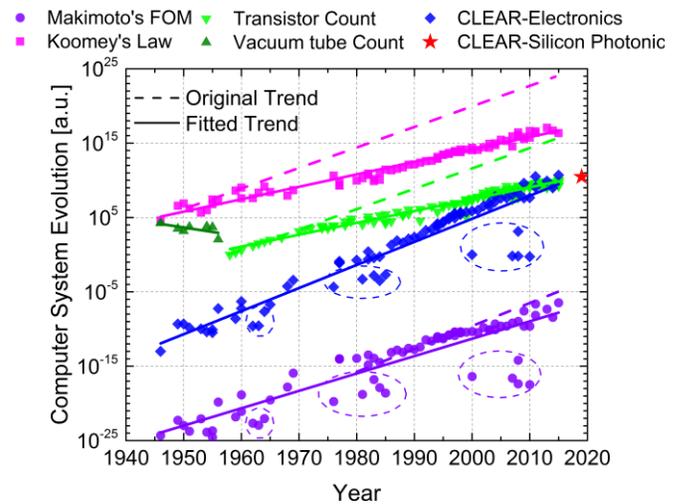

Fig. 1. Computer System Evolution Trends. The solid lines denote the actual growth trend of corresponding FOM by fitting the computer data from 1946 to 2016, and the dashed straight lines represent the original trend that each FOM model predicted in the beginning. The y-axis with arbitrary unit has different physical meanings for different FOMs and shows the relative growth rate of each prediction model.

III. HOLISTIC FIGURE-OF-MERIT FOR COMPUTE SYSTEMS

The analysis and comparison above shows that an appropriate FOM capable of accurately post- and predictions of the evolution of compute systems needs to include a holistic set of factors that incorporate physical and even economic constrains. Here, we introduce a universal and multi-hierarchical FOM based-on Capability-to-Latency-Energy-Amount-Resistance (CLEAR, Eqn. 1). At a higher level, this performance-to-cost model connects fundamental constrains such as entropy-per-bit and bit-error-rate (BER) with economic replacement and adoption models. It is adaptable to be applied to different hardware hierarchy levels such as compute systems, links, or individual devices.

$$\text{CLEAR} = \frac{\text{Performance}}{\text{Cost}} = \frac{\text{Capability}}{\text{Latency} \times \text{Energy} \times \text{Amount} \times \text{Resistance}} \quad (1)$$

The individual factors in CLEAR are defined based on the hierarchy levels it applies to; for instance, at the compute system level, CLEAR breaks down as follows: the capability (C) is the system performance given by million-instructions-per-second (MIPS); the minimum latency (L) relates to the clock frequency and is limited by the temporal window between two adjacent clock cycles; the energy efficiency (E) represent energy cost for operating each bit in the units of joule-per-bit; the amount (A) represents the spatial volume of the system and is a function of the process dimensionality; the resistance (R) quantifies the economic resistance against a new technology adoption. It is an economic model based on the Boston Consulting Group (BCG) experience curve which explains the relation between the cumulative production and the unit cost. We derive the linear relation between the log scale of unit price and time, and this relation could be confirmed by the historical data of transistor [30]. We note that while the metric MIPS as a measure of performance is being replaced by metrics such as floating point operations (FLOPS) due to its susceptibility to the underlying instruction set, in this work CLEAR is applied to many historical processors for which other performance metrics are not available under known benchmarking suites (for example SPEC or LINPAC). Towards making MIPS a representative performance metric however, we weighted (i.e. multiplied) each instruction by the length of its representation, thus giving the relative time for completion of the execution.

*A. Computer System Evolution Trend*

The results show that the five-factor FOM CLEAR is able to accurately track the entire computer system evolution, which displays a constant growth rate enabled by the holistic FOM (Fig. 1). Moreover, the actual observed evolution rate is consistently held at 2x for every 12-months. Comparing CLEAR to Moore's Law and Makimoto's FOM, we find that the FOM which incorporates additional relevant factors will only deviate from the actual trend later. With five related factors, CLEAR does not deviate from the 2x/year trend even as different technologies supersede each other (vacuum tubes, transistors) Testing emerging optoelectronic-based compute engines we find that such technologies indeed appear to continue the evolutionary 2x/year development trend (red data, Fig. 1) [31].

In addition, applying this multi-factor FOMs we are able to classify compute systems by their relative position to the 2x/year trend line (Fig. 1). For instance the additional overhead on (i.e. physical space, parallelism, heat removal, low economy-of-scale, manufacturing costs) of supercomputers show their inferior CLEAR relative to all other computer types, despite their higher performance (dashed circles, Fig. 1). The high parallelism of multi-core technologies used in supercomputers is challenged by compute-to-energy returns described by Amdahl's law [32]. We observe that while supercomputers deliver peta-FLOP performance, they entire infrastructure resembles that of computers 5-30 years back thus questioning future scale-up.

### B. Computational and Energy Efficiency Tradeoff

As mentioned before, supercomputers are trading their energy efficiency for extra computing power. However, their actual computational efficiency, which represents the amount of data capacity that a system is able to handle per unit latency, area and price, is slowly saturating. Computational efficiency is a key quantifier describing capability improvement with respect to i) processing data as a function of energy consumption, and ii) to achieve higher computing efficiency per unit resources (Fig. 2). Conventional electronic compute systems are facing an efficiency wall around $10^4$ instructions-per-joule and computational efficiency <1 Petabyte ($10^{15}$) per unit delay, volume and economic cost (in the units of second, $mm^3$ and \$). For electronics this can be understood as an increased 'resistance'; despite continued transistor scaling, fabrication cost per unit and clock speed are strained due to process yield control at the atomistic scale and electrical RC interconnect delay, respectively. Moreover, parasitic 'dark silicon' and heat dissipation reduce the design window, thus preventing an increase number of cores in a CPU. This dichotomy limits electronic compute systems thus 'boxing' in electronics (black data, Fig 2).

Interestingly, the state-of-the-art silicon photonics roadmap for manufacturing demonstrated by IBM outperforms electronics in both energy- and computational-efficiency by several orders of magnitude (red data, Fig. 2). This long-term prediction illustrates that the energy efficiency of silicon photonics might be able to break through the electronic energy efficiency wall to reach the pico- to femtojoule level and multiple orders of magnitude higher bandwidth that electronics is unable match due to capacitance. Main advantages of photonics are found in a) the low attenuation loss during signal propagation, b) the independence from electrical capacitance, c) compatibility with WDM enabled by the high Finesse of resonators on-chip, and d) analogue computing options to deploy more information in a signal such as utilizing amplitude, phase, and polarization simultaneously leading to high information-per-entropy density [33]. Despite known drawbacks such as the DLL challenge scaling of photonics, and low LMIs requiring a high control voltage for light manipulation, silicon photonics is proven to have high computational efficiency. As such is already being adapted to emerging compute engines such as neuromorphic computing, reconfigurable optical networks, and metamaterial-based computing [34-35]. Comparing Fig. 2 with Fig. 1, we conclude that photonics has higher compute potential than traditional electronics in terms of both energy and compute efficiency.

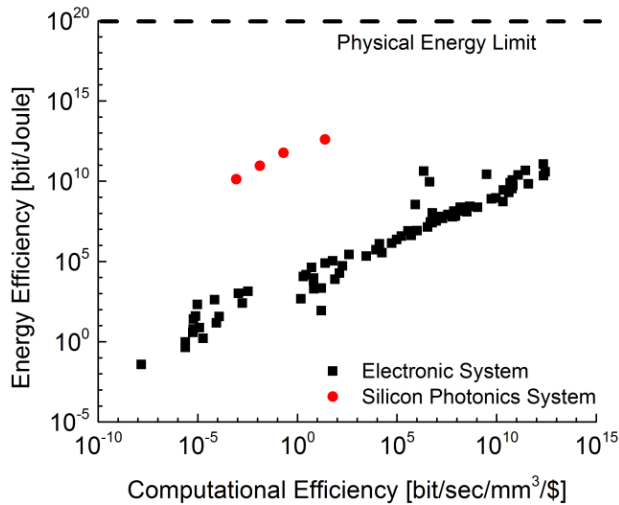

Fig. 2. Energy efficiency and computational efficiency trade-off. The black solid dots denote the computer system data points from Fig.1 and the red circles represent the prediction for on-chip silicon photonics made by IBM [31]. The x-axis represents the latency, volume and fabrication cost to generate one bit of information, while the y-axis shows the energy cost of it. The physical energy limit at $10^{20}$ bit per Joule level is derived based on the Landauer's principle which bounds the theoretical energy consumption of computations.

IV. OTHER HIERARCHIES APPLICATIONS OF CLEAR

Above we showed that CLEAR is an appropriate FOM to trace the information processing capability of a compute system. Next we show that this metric can also be used for device- and link level comparisons provided small amendments are made, thus making CLEAR multi-hierarchical. Our discussion includes a comparison between traditional electronics, emerging photonics, plasmonics, and hybrid photonic-plasmonics among device, interconnect and network levels.

*A. CLEAR Comparison at Device Level*

In order to amend CLEAR for device-level usage, a few adjustments need to be made to capture the characteristics of device components. Here CLEAR becomes Capability-to-Length-Energy-AREA-Ratio, which breaks-down as follows: i) the device operating frequency is the capability (C); ii) the scaling efficiency which is the reciprocal of the critical scaling length (L) of the device describes the interaction length to provide functionality; iii) the energy consumption (E) of the energy 'cost' per bit is the reciprocal of the energy efficiency; iv) the on-chip footprint, or area (A), and v) the economic resistance (R) in units of dollars ($) is the reciprocal of the device cost efficiency. Here the critical scaling length in the denominator does not conflict with the area factor, but indicates the scaling level or ability of the device to deliver functionality given its length. For instance, the critical scaling length of the CMOS transistor is the length of its logic gate, which controls the ON/OFF states. For photonic and plasmonic devices, it can be regarded as the ring diameter and the side length of the active layer respectively.

We represent the device-CLEAR results as five merit factors in a radar plot (Fig. 3). Note, each factor is represented in such a way that the larger the colored area in Fig. 3 the higher the CLEAR FOM of the device technology. Moreover, some of the factors of the device-CLEAR have fundamental physical constrains that prevent them for further growth despite the technology. For example, the energy efficiency of the device is ultimately limited by the Landauer's principle ($E_{min} \geq k_B T \ln(2)$), which restricts the minimum energy consumption to erase a bit of information to 2.87 zeptojoule at room temperature (T = 300K) [39]. Given this device energy limit, Margolus–Levitin theorem set a cap for the maximum operating frequency of the device. Based on the fundamental limit of quantum computing, a device with the amount of $E$ energy requires at least $h/4E$ of time to transfer from one state to the other resulting in over 16 THz for energy levels approaching the Landauer's limit [40]. When approaching the quantum limit for data communication, the device's critical length would be scaled down to the dimension of about 1.5 nm based on the Heisenberg uncertainly Principle $(x_{min} \geq \hbar/\Delta p = \hbar/\sqrt{2mE_{min}} = \hbar/\sqrt{2mk_B T \ln 2} = 1.5 nm)$.

With all five CLEAR axis in Figure 3 normalized to their respective physical limits, we compare the device performance of four different technologies. Results show, that the overall performance a single transistor is extremely high benefitting from small RC delay times, compact sizes due to the nanometer-small wave function of electrons, and cost-efficiency from economic learning curves over 5-decades. However, emerging technologies, such as photonics and plasmonics, are gradually catching up with the electronic devices, and their hybrid combination may exceed the transistor at operating frequencies that are limited by the device RC delay (R = resistance, C = Capacitance) due to the sub-diffraction limited active region below 100 nm dimension and low insertion losses due to modal silicon-plasmonics hybridization. However, the burgeoning optics-based technologies are still suffering from the immature fabrication process and incomplete economic industry system leading to high cost, low scaling- and low area efficiency. Area-wise, those emerging technologies are unable to surpass the energy efficiency of a single transistor. Nevertheless, this situation will significantly change when applying CLEAR to the interconnect-level comparison discussed below. Because of the low light attenuation of the passive silicon-on-insulator (SOI) waveguide, the photonic and hybrid photonic-plasmonic interconnect show a high CLEAR especially for chip-scale signaling distances; while electronic and plasmonic links require extra overhead due to high transmission losses.

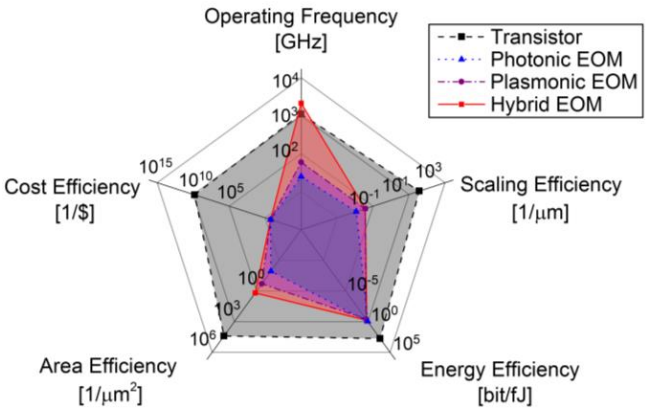

Fig. 3. The CLEAR comparison at device level. Each axis of the radar plot represents one factor of the device-CLEAR and is scaled to the actual physical limit of each factor. Four devices compared from different technology options are: 1) the conventional CMOS transistor at 14 nm process; 2) the photonic microdisk silicon modulator [36]; 3) the MOS field effect plasmonic modulator [37]; and 4) the photonic plasmonic hybrid ITO modulator [38]. The colored area of each device also demonstrates the relative CLEAR value of each device.

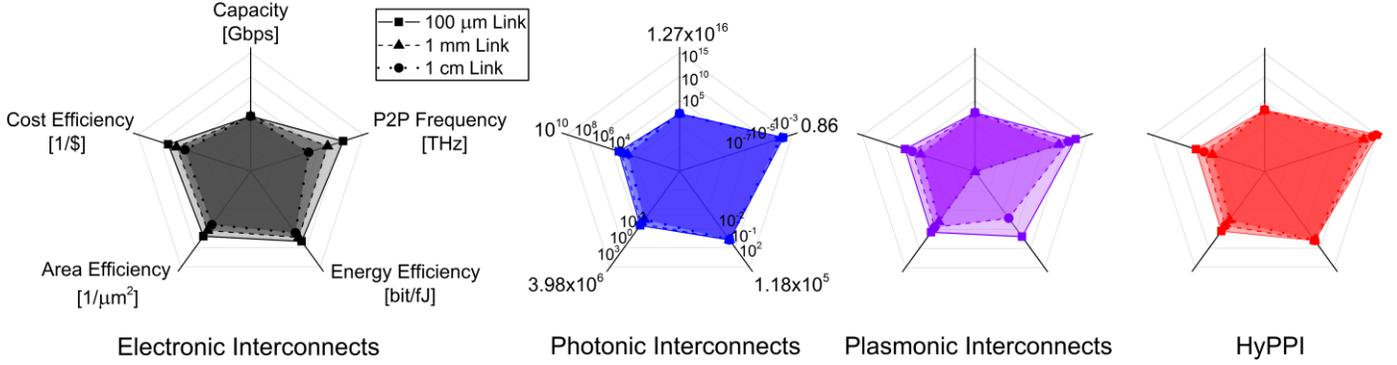

Fig. 4. Link-CLEAR breakdown comparison among 1) electronic, 2) photonic, 3) plasmonic and 4) hybrid photonic-plasmonic interconnects. The solid lines covered areas with square nodes represent the link-CLEAR for 100 μm length interconnects, and the dashed lines with triangle nodes and the dotted lines with circular nodes represent the 1 mm and 1 cm length interconnects respectively. All five link-CLEAR are shown at the same axis of the four radar plots and scaled into the same range constrained by the physical limit. The numbers at the end of each axis of the second radar plot show the actual physical limits of each factor. All of the four radar plots share the same axis titles and ticks, and certain titles and tricks are omitted for conciseness.

*B. CLEAR Comparison at Link Level*

Next, we apply CLEAR to the link-level, and again consider the same four technology options as for the device level. The link is defined as a transmission line between a sender and a receiver, and only devices relevant to its own technology are considered. For instance, the hybrid photonic-plasmonic interconnect (HyPPI) uses a nano plasmonic laser as the light source, a hybrid-plasmon ITO-based modulator that we compared in the device-CLEAR section and lastly detect the light by a plasmonic photodetector. Indeed the CLEAR HyPPI shows a high on-chip performance, which includes high energy efficiency, a low point-to-point latency, and a high throughput [5]. However, these metrics are only able to capture a single performance aspect, while CLEAR combines the multitude of factors into a single value, thus allowing for an objective comparison. To adjust the CLEAR model for the link-level comparison, i) the capacity (C) is defined as the maximum data rate of the interconnect channel that can deliver after compensating the communication noises through the channel. It describes how fast the interconnect is able to transmit from the sender to its receiver while taking into account the channel crosstalk, device bandwidth, transmission loss, detector sensitivity and minimum output current requirement. ii) The point-to-point (P2P) latency (L) is defined as the time-of-flight of a bit of information propagating from the source to the photodetector. iii) The energy consumption (E) represent the energy sum consumed by all the devices on the link including all the related drivers and accessorial components [5]. iv) The area (A) is given by the sum of all devices on-chip footprint for the interconnect. v) Finally, the economic resistance (R) uses the similar model used in the device-CLEAR section.

Here, we compare the link-CLEAR among the four aforementioned technologies for three different link lengths, (100 μm, 1 mm, 1 cm) to study the signal length dependent performance change (Fig. 4). Note, the chip-scale (i.e. die size) is about 1 cm. All five axis of Fig. 4 are normalized to the physical limit similar to the device analysis. For instance, the capacity of the link is restricted by the Bremermann's Limit, which is derived from the mass-energy equivalency and the Heisenberg uncertainty principle giving a maximum bit rate per unit mass of the system [41]. Assuming the link only contains of smallest devices of 1.5 nm each for the sender and receiver, as the 'ultimate' case, this system is still able to provide over $10^{16}$ bps data rate as shown on the capacity axis. The physical limit for the P2P latency can be basically regarded as the light propagation time through the link. The maximum P2P frequency for a 100 μm is approximately 1 THz, and for 1 mm and 1 cm distance, the frequency limit is one and two orders of magnitude lower. Energy-wise the Landauer's limit applies to the device-CLEAR section, the energy efficiency of this ultimate link is half of the device level energy efficiency since that bit of information has been manipulated twice. Moreover, the area efficiency limit of the link level is also half of that of the device limit level. However, for the economic part, there is no actual physical limit since the fabrication processes we are still being scaling. Thus, here we take the cost efficiency axis limit to be $10^{10}$ but not that it still might improve with time.

For all four interconnect options we find that there is significant room for development. For the P2P frequencies, all three optical interconnects show higher performance compared to electronic interconnects due to low RC delay times, especially for HyPPI, which uses passive low-loss SOI waveguides as for signal propagation and LMI-enhanced active optoelectronics. Although HyPPI is able to deliver about 10-100 times higher capacity compared to the other options, it still falls short several orders of magnitude before we approaching ultimate limits. In addition, the energy efficiency of the photonic and HyPPI interconnect scale relatively independent with link length, due to the low-loss signal propagation in the waveguides even at high speeds, while the electronic and plasmonic interconnects have increased overhead penalties with link lengths. Note, the area efficiency of the four interconnects is similar to their capacity model, which is determined by the slowest speed of the device on the link, the signal-to-noise ratio of the entire link and the point-to-point latency, that has minimal or no penalties for up scaling within the distance limits of the chip range. To be more specific, the capacity of channels under different length are the same since they all have a fixed bandwidth and the light propagation time does not affect the overall latency due to the short distance. For the area efficiency, the area increases with the interconnect length linearly which does not require extra on-chip footprint penalty. Lastly, the electronic interconnect shows a ~10x higher cost efficiency highlighting the development effort of the semiconductor industry. Despite this, all interconnect options have significant room for further improvement until fundamental physical limits are reached. To distinguish the improvement of different interconnect technologies with development time, we adapted the time-model into each factor of CLEAR and showed the technology evolution over the years in another our recent work [42]. Breaking down CLEAR for both the device and link-level into its five components reveals insights into the various performance capabilities of these technologies. While electronics is clearly performing well at the device level, latency and energy limitations for optical options become apparent at the link level. Based on these detailed performance merits, the link technology options can be used as a basic indicator for multi-technology network-on-chips (NoC). In addition, reconfigurable networks are conceivable, which allow the network select between varieties of link options depending on the application demand.

C. *CLEAR Comparison at Network Level*

We next apply CLEAR to the network level and compare the different link technology options for a 16×16 Mesh network-on-chip (NoC). Adjusting the definitions of the individual factors in CLEAR applicable to a network gives:

$$\frac{(\sum_{i=1}^{all\ links} C_i)/N}{Latency\ (clks) \times Energy/bit \times Area \times Resistance} \quad (2)$$

$C_i$ is the bandwidth capacity of link $i$, and $N$ is the number of nodes. The capacity (C) is therefore the aggregate link capacity averaged across all nodes. Latency (L) is the average number of clock cycles incurred by the flits as they traverse from their source node to their destination node. The Resistance factor (R) is an estimated economic cost based on the wafer costs and the area occupied by the electronic and photonic components on their respective dies. Note that we adopt a NoC that uses electronic routers and point-to-point links between the routers.

In this evaluation, we study the effect of using different technologies for these point-to-point links.

**NoC parameters:** The following set of NoC parameters are used in our evaluation; a 16x16 mesh network with 1 mm inter-core spacing. All links are rated at 50 Gb/s irrespective of the technology option selected. We use 32-bit flits, and correspondingly, the electronic links are 32-bits wide running at 1.5625 GHz. Note that Plasmonics and HyPPI are capable of capacities beyond 50 Gb/s; however, contemporary electronics required for the link drivers and SERDES circuits are capable of only ~50 GHz, thus limiting link speeds. Photonics, however, is capped at 25 Gb/s and thus needs two wavelength channels [4]. For the energy and area estimates, we used the DSENT tool for an analysis of the links and routers, adopting the 11 nm technology node [43]. For HyPPI, we modified DSENT based on previously published component parameters [5]. For Plasmonics we repeated the link every 100 µm, due to its losses.

The latency for electronic links is 1 clock cycle. For optical links, however, an additional clock cycle is added in order to account for the O/E conversion at the receiver, because the routers are electronic. The link propagation delay is bounded within the clock cycle of the 1.5625 GHz router clock used. Thus, all optical links exhibit 2-clock cycles latency. The router pipeline latency is three clock cycles.

**Evaluation Methodology:** We use a synthetic traffic statistics to model input traffic, based on Soteriou et. al [44].

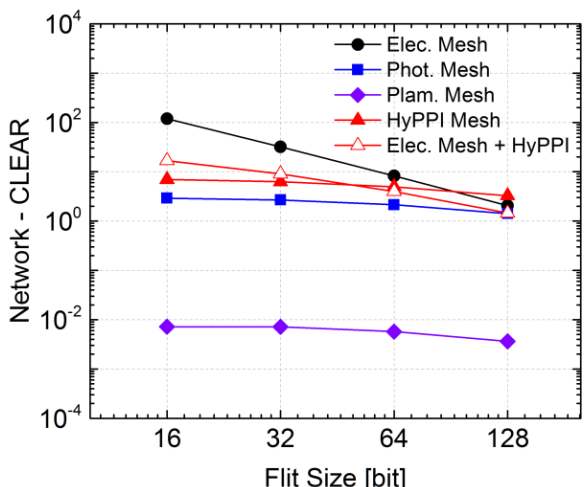

Fig. 5. Comparison of network-level CLEAR for a 16x16 Mesh NoC using different technologies for its individual links. NoCs compared from left to right are 1) electronic mesh; 2) photonic mesh; 3) plasmonic mesh; 4) HyPPI mesh and 5) electronic mesh integrated with HyPPI links. The CLEAR units are: Gb/s for C, clock cycles for L, pJ/bit for E, mm$^2$ for A, and $ cost for R.

We then estimate the activity on each link in the mesh network. Subsequently, we compute the total dynamic energy per bit accounting for all links and routers based on the injection rate at each link. The throttled laser model from DSENT is adopted, and thus the laser power is accounted for as part of the dynamic energy consumption. The total area is obtained by summing the individual component areas obtained from DSENT, which includes routers, links, drivers, and SERDES.

**Results:** .As expected, the higher economic costs of all the optical link options reduces the CLEAR value, indicating that an electronic mesh is the most viable option at this point of time. However, it will be surpassed by HyPPI mesh for increased flit size (128 and larger) basically due to less energy and area efficiency. Nevertheless, HyPPI shows a significant advantage over conventional silicon photonics, because of its lower area and energy requirements. Furthermore, plasmonics is the least suitable options, due to higher energy costs given the assumed 1 millimeter core-to-core link lengths. As photonic wafer costs reduce with economic learning curves, we expect HyPPI to eventually become viable for NoC interconnect. On the path towards HyPPI-enabled NoCs, we envision that the industry will witness a gradual transition from electronics links to HyPPI. For example, we repeated our experiment for the same 16x16 electronic mesh, augmented with HyPPI "express links". These HyPPI links are used for connecting cores in the horizontal direction that are 3 hops apart in the electronic mesh. Thus, each row in this network has 5 additional HyPPI links to aid the electronic links. This results in energy efficiency enhancements and low-latency advantages of HyPPI for long-range traffic, while leveraging cheaper electronic links for shorter communications. Fig. 5 shows that the CLEAR value for this HyPPI-augmented electronic network is higher, and is thus a good first step towards fully optical NoCs.

## V. CONCLUSION

We introduce a novel Figure-of-Merit, CLEAR, incorporating a holistic set of performance parameters. This FOM is universal since it covers both physical and economic factors known to-date that contribute to the evolution of computer systems. As such it is applicable to a variety of technologies to include electrical and optical options. CLEAR reveals the constant growth rate of compute system across different technology cycles, and can therefore be used for post- and predictions in technology development. Furthermore it bears universality as it can be applied to conduct device/link/network/system comparisons. We show that electronics is competitive at the device level. At the link level, only the separation of active and passive functionality between plasmonics and photonics leads to performance that surpasses electronics. Testing a 256-core mesh network we show that CLEAR reveals the actual network performance indication hybrid photonic-plasmon interconnects to become competitive for flit size above 128 bits. Founded on fundamental physical principles, this FOM has the potential to become the next Moore's law for both data processing and computing.

ACKNOWLEDGMENT

This work was supported by the Air Force Office of Scientific Research (AFORS) with award number FA9550-15-1-0447.